\documentclass[useAMS,usenatbib]{mn2e}

\usepackage{natbib}
\usepackage{graphicx}
\usepackage{multicol}
\usepackage{longtable}
\usepackage{amssymb}
\usepackage[width=1.0\textwidth]{caption}
\usepackage[usenames,dvipsnames]{color}
\RequirePackage[pdftex, plainpages = false, pdfpagelabels]{hyperref}
\definecolor{Red}{rgb}{1.0,0,0}

\newcommand{\etal}    {{\it et al}}                          
\newcommand{\SL}  [2] {$^{#1}$#2}                  
\newcommand{\SLP} [3]{$^{#1}$#2$^{\rm{#3}}$}            
\newcommand{\SLJ} [3]{$^{#1}$#2$_{\rm{#3}}$}            
\newcommand{\SLPJ}[4]{$^{#1}$#2$^{\rm{#3}}_{_{#4}}$}       
\newcommand{\nlo} [3]{#1#2$^{#3}$}                 
\newcommand{\po}  [2]{\overline{#1}\rm{#2}}        

\newcommand{\AS}    {Autostructure\index{Autostructure}}

\newcommand{\II}      {~{\sc ii}}
\newcommand{\III}     {~{\sc iii}}

\newcommand{\otp}     {O$^{2+}$}
\newcommand{\ntp}     {N$^{2+}$}
\newcommand{\netp}    {Ne$^{2+}$}
\newcommand{\op}      {O$^+$}
\newcommand{\CR}      {}

\title[{\rm O}\II\ Recombination] {Recombination coefficients for O\II\ lines in nebular conditions}
\author[P.J. Storey, Taha Sochi \& Robert Bastin]
{P.J. Storey$^{1}$, Taha Sochi$^{1}$\thanks{E-mail: t.sochi@ucl.ac.uk}, Robert Bastin$^{2}$ \\
$^{1}$Department of Physics and Astronomy, University College London, Gower Street, London, WC1E
6BT, UK \\
$^{2}$Surbiton High School, Kingston upon Thames, Greater London, Surrey, KT1 2JT, UK}

\begin{document}

\date{Accepted 2017 May 11. Received 2017 May 10; in original form 2017 March 28}

\maketitle

\label{firstpage}
\begin{abstract}
\noindent We present the results of a calculation of recombination coefficients for \otp + e$^-$
using an intermediate coupling treatment that fully accounts for the dependence of the distribution
of population among the ground levels of \otp\ on electron density and temperature. The calculation
is extended down to low electron temperatures where dielectronic recombination arising from Rydberg
states converging on the \otp\ ground levels is an important process. The data, which consist of
emission coefficients for 8889 recombination lines and recombination coefficients for the ground
and metastable states of \op\ are in Cases A, B and C, and are organised as a function of the
electron temperature and number density, as well as wavelength. An interactive fortran 77 data
server is also provided as an accessory for mining the line emission coefficients and obtaining
Lagrange interpolated values for any choice of the two variables between the explicitly provided
values for any set of wavelengths. Some illustrations of the application of the new data to nebular
observations are also provided.
\end{abstract}

\section{Introduction} \label{Introduction}

O\II\ recombination lines have been observed in the spectra of many planetary nebulae and H\II\
regions \citep{PeimbertSP1993, LiuSBC1995, BaldwinVVFMe2000, GarnettD2001, WessonLB2003,
EstebanPGRPR2004, WessonLB2005, TessiG2005, PeimbertP2005, LiuBZBS2006, EstebanBPGPM2009,
Delgado2010, PeimbertP2013, PeimbertPIRP2014}, and hence many of these lines (e.g.
$\lambda\lambda$4076, 4089 and the V1 multiplet) have been used \citep{LiuLBL2004, TsamisBLSD2004,
WessonLB2005, BastinThesis2006, BastinS2006, WangL2007, McnabbFLBS2013} as diagnostic tools for
probing the physical conditions of the emitting regions.

Regarding previous theoretical and computational work on \op\ recombination processes,
\citet{NussbaumerS1984} calculated the effective dielectronic recombination coefficients for a
number of selected O\II\ recombination lines in an $LS$-coupling approximation.
\citet{PequignotPB1991} calculated total and effective radiative recombination coefficients for
several important O\II\ lines and provided them in the form of fitting formulae valid over certain
temperature ranges. Similar calculations were carried out by \citet{Smits1991} for selected lines
where the calculations were based on an $LS$-coupling scheme. The work of \citet{NussbaumerS1984}
was extended by \citet{Storey1994} who computed effective recombination coefficients for a number
of O\II\ lines based on a full $LS$-coupled \op\ model atom at electron temperatures and number
densities relevant to planetary nebulae. \citet{LiuSBC1995} provided O\II\ radiative recombination
coefficients for the 3d-3p and 4f-3d transitions based on the results of \citet{Storey1994} but
with transformation of radiative data to an intermediate coupling scheme limited to these
transitions. All of the above calculations assumed that the three levels of the \otp\ \SL3P ground
term are populated according to their statistical weights, which is not usually the case in the conditions
typically found in photoionised nebulae. {\CR In general, for any ionic level, if the ambient
density is below the critical density for that level, its population will depart from a Boltzmann
distribution and the population of all states formed by recombination on to that level will be
affected. The result is that for recombination on to an ion with more than one level in the ground
term, such as \otp, the relative intensities of lines within all recombination multiplets will, in
general, show some density dependence.}

\citet{RuizPPE2003} investigated this effect empirically by considering the observed variation of
the relative intensities of the lines of O\II\ multiplet V1 as a function of the electron number
density derived from forbidden lines, mainly [Cl\III]. This approach was further refined by
\cite{PeimbertPR2005} and \cite{PeimbertP2005}. We compare these authors' results with our
theoretical work in a subsequent section. \citet{BastinThesis2006} and \citet{BastinS2006} revised
the calculation of \cite{Storey1994} by describing the whole recombination process in intermediate
coupling and including the distribution of population among the \otp\ levels. They used an R-matrix
scattering code \citep{BerringtonEN1995} to compute oscillator strengths and photoionisation
cross-sections, and hence recombination coefficients, for the low lying states of \op.
\citet{McnabbFLBS2013} presented various O\II\ diagnostic line ratios derived from the theoretical
results of \cite{BastinThesis2006} and used them to analyse a large number of planetary nebula and
H\II\ region spectra.

In the present paper we use the techniques used by \cite{BastinThesis2006} and \cite{BastinS2006},
and later also used by \cite{FangSL2011} for the recombination of \ntp, but with some modifications
and improvements as described below. The results presented here differ, in some cases
significantly, from those obtained by \cite{BastinThesis2006}. We return to this point in the
relevant subsequent sections. We provide the resulting data in a text file containing a list of
transitions between the lower states of \op\ and emission coefficients for these lines as a
function of electron temperature and density and Case (A, B and C which will be explained later).
We also provide a file containing recombination coefficients for the ground and metastable states
of \op\ as functions of the same quantities. We also provide an interactive data server, in the
form of a fortran 77 program, for mining the emission coefficient data and obtaining interpolated
results for densities and temperatures between the explicitly computed values. More details will be
given in the forthcoming sections.

As discussed briefly in \citet{StoreySSS82015}, the current paper comes as part of a series of
papers by the authors \citep{StoreySSS22013, StoreySSS42014, StoreySB2014, StoreySSS72015,
StoreySSS82015} aimed at providing the means to investigate the long-standing puzzle of the
inconsistency between the results of elemental abundance and electron temperature as derived from
the optical recombination lines (ORL) with those derived from the collisionally-excited lines
(CEL). Several possible explanations for the discrepancy have been proposed including the proposal
that the recombination lines are formed in regions of much lower temperature than implied by the
CEL diagnostics. We therefore extend the calculation of recombination coefficients down to much
lower electron temperatures (100~K) than is usually the case.

It has also been suggested that the free electron energy distribution may not be best described by
a Maxwell-Boltzmann distribution \citep{NichollsDS2012, NichollsDSKP2013}, so we plan to publish a
further paper where recombination coefficients have been calculated with the free electron energies
distributed according to the $\kappa$ distribution suggested by theses authors. We note, however,
that the non-Maxwellian model at present has little theoretical \citep{FerlandHODP2016} or
observational \citep{StoreySSS42014, ZhangZL2016} support.

The plan of the present paper is that in section~\ref{Computation} we give details of the
calculation of the atomic parameters needed to calculate the population structure, including
details of a new R-matrix calculation of radiative properties. In section~\ref{Data} we give
details about the structure and contents of the data files and provide general instructions and
explanations about how the data should be explored and used. Section~\ref{ExampleResults} presents
some comparisons between theory and observation and the paper is concluded in
section~\ref{Conclusions} with general conclusions.

\section{Atomic parameters for \op}\label{Computation}

 In this section we summarise the main features of the past and present
calculations. We follow the methods used by \cite{Storey1994}, \cite{BastinThesis2006},
\cite{BastinS2006} and \cite{FangSL2011} to compute the atomic parameters required to compute
populations of the states of \op\ due to recombination processes. We refer the reader to those
references for a fuller explanation.

\subsection{The \op\ Term Scheme}\label{TermScheme}

\begin{figure}
\centering{}
\includegraphics[scale=0.53]{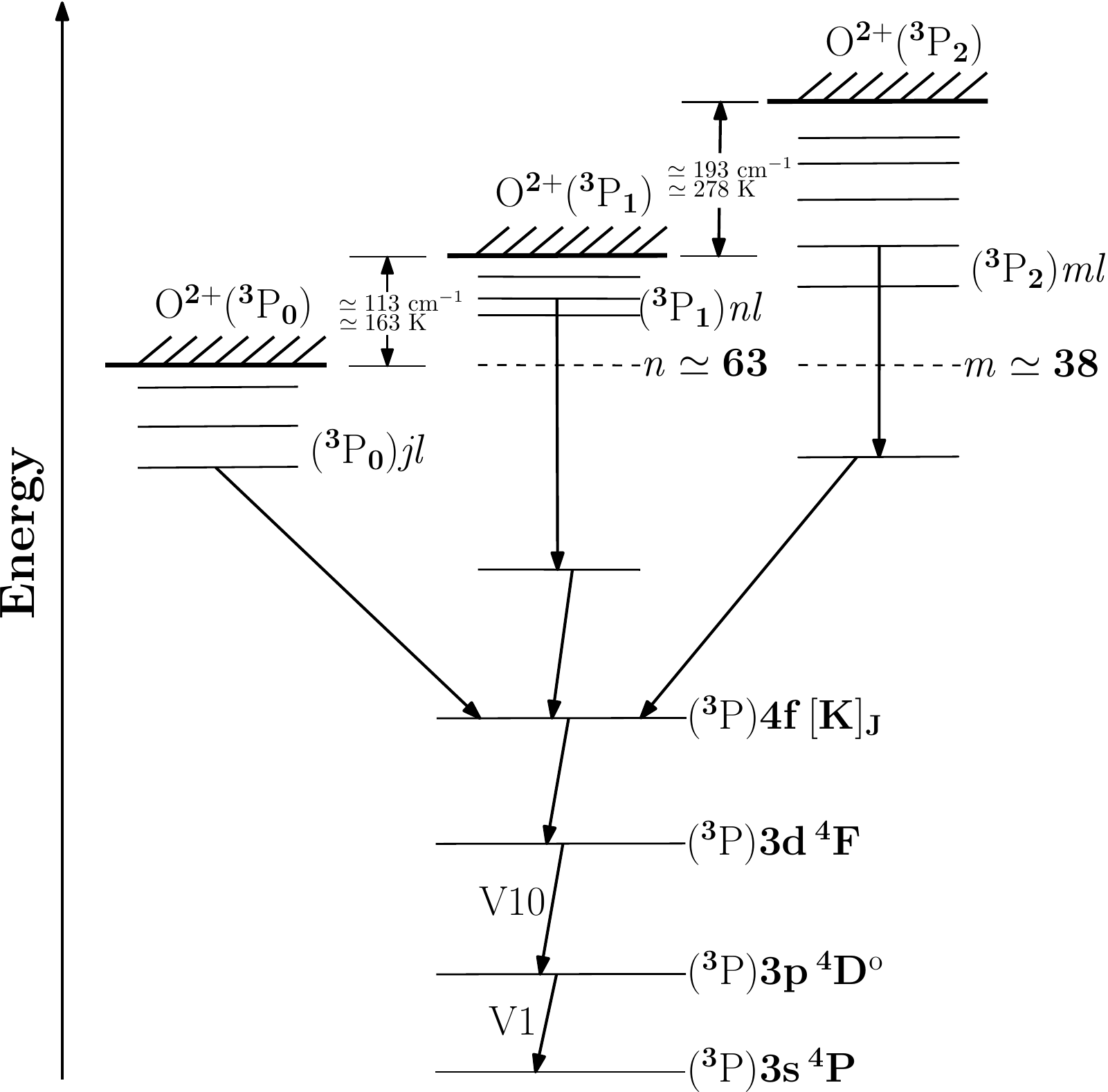}
\caption{Schematic diagram of low-lying energy states of O$^+$  with some low-temperature
dielectronic recombination transitions between fine-structure autoionising levels below the three
lowest ionisation thresholds of O$^{2+}$: $^3$P$_0$, $^3$P$_1$ and $^3$P$_2$. Some multiplets
arising from transitions between low-lying terms are also depicted schematically. The figure is
qualitatively informative but is not drawn to scale.} \label{DRfig3}
\end{figure}

The principal series of \op\ is 2s$^2$2p$^2$($^3$P)$nl$ which gives rise to doublet and quartet
terms. Embedded within this series are the 2s$^2$2p$^2$($^1$D)$nl$, 2s$^2$2p$^2$($^1$S)$nl$, and
2s~2p$^3$($^5$S$^{\rm o}$)$nl$ series and terms of the 2s~2p$^4$ configuration. Some of the members
of these additional configurations lie above the first ionisation limit and can be a source of
population {\it via} dielectronic recombination. In the calculation of \cite{Storey1994} the states
and radiative transitions of \op\ were described in $LS$-coupling but in this approximation several
important processes could not be incorporated. To remedy this shortcoming, \cite{BastinThesis2006}
described the states of \op\ in a $J_cj;J$ coupling scheme, where $J_c$ is the total angular
momentum quantum number of the \otp\ state, $j$ is that of the valence electron and $J$ is the
total for the \op\ state. This enabled two significant extensions to the work of \cite{Storey1994}:
firstly, the populations of the $^3$P$_{J_c}$ levels of the \otp\ recombining ion could be
explicitly accounted for, and secondly the three series of Rydberg states converging on the
$^3$P$_{J_c}$ levels could be treated separately. Figure~\ref{DRfig3} illustrates these states
schematically. These Rydberg states are a channel for a dielectronic recombination process that
operates at very low temperatures, below approximately 1000~K. In their calculation of \ntp\
recombination, \cite{FangSL2011} used the methods described by \cite{BastinThesis2006} but modified
the coupling scheme describing the \op\ states to a $J_cl(K)s;J$ scheme, where $l$ and $s=1/2$ are
the orbital and spin angular momenta for the valence electron. In this scheme $J_c$ is coupled with
$l$ to give the quantum number $K$, which in turn is coupled with $s$ to give the total angular
momentum quantum number $J$. This scheme has the advantage that it is the natural scheme for the
high-$l$ Rydberg states which appear in energetically close pairs characterised by $J=K\pm 1/2$. It
is also the scheme used to describe the continuum states in the R-matrix electron scattering codes
described below in Section~\ref{Rmatrix}. The experimental results for the energy levels of \op\
given by \cite{MartinKM1993} were used here to assign $K$ and $J$ quantum numbers to \op\ states
with $l>2$. For states of lower $l$ a transition to $LS$ coupling takes place which is fully
accounted for in the intermediate coupling R-matrix treatment of the bound states described below.

\subsection{\op\ Populations}

\citet{FangSL2011} have described the calculation of the departure coefficients $b(J_c nl(K)s;J)$
defined by:
\begin{equation}
 \frac{N(J_c nl(K)s;J)}{N_e N_+(S_cL_c;J_c)} =  \left( \frac{N(J_c nl(K)s;J)}{N_e N_+(S_cL_c;J_c)}\right)_S b(J_c nl(K)s;J)
\end{equation}
where $N(J_c nl(K)s;J), N_+(S_cL_c;J_c)$ and $N_e$ are the number densities of the \op\ state $J_c
nl(K)s;J$, the \otp\ state $S_cL_c;J_c$ and free electrons respectively and the subscript $S$
signifies the ratio of the number densities given by the Saha-Boltzmann equation. There are three
stages to the calculation as described originally by \citet{Storey1994}. To summarise, the first
stage involves a calculation of $b(J_c n)$ for all $n \le 1000$ for the $J_c=0,1,2$ of the $^3$P
ground term of \otp. In addition to the collisional-radiative processes described by
\citet{HummerS1987} for hydrogen-like systems, autoionisation probabilities to and dielectronic
capture probabilities from the available continua, averaged over $l$ for each $n$, are
incorporated. For $n>1000$ we assume $b(J_cn)=1$. In the second stage, departure coefficients
$b(J_c nl(K)s;J)$ are calculated, still in a hydrogen-like approximation for all $n<n_l$, where
$n_l$ is the boundary above which $l$-changing collisions are sufficiently rapid that $b(nl)=b(n)$
is a good approximation. In the first two stages there are necessarily no radiative or collisional
processes linking states of different $J_c$. The third and final stage deals with the departure
coefficients for the states with $n<n_d$, where $n_d$ is the boundary below which only radiative
processes determine the populations, i.e. where collisional processes can be neglected.

\subsection{R-matrix Calculation for \otp + $e^-$}\label{Rmatrix}

For the low-lying states of \op, the recombination coefficients were computed directly by
integrating the photoionisation cross-sections, as obtained from an \otp + $e^-$ R-matrix
calculation in an intermediate coupling scheme using the Breit-Pauli approximation. Cross-sections
were obtained for all the states with principal quantum numbers $n\le 10$ and orbital angular
momentum quantum numbers $l\le 4$ in the principal series of \op\ plus all other states embedded
among these. The calculation also provides bound state energies and oscillator strengths for use in
the calculation of the \op\ level populations. The main codes used for the calculation are
\AS\footnote{{See Badnell: Autostructure write-up on WWW. URL:
\url{amdpp.phys.strath.ac.uk/autos/ver/WRITEUP}.}} \citep{EissnerJN1974, NussbaumerS1978, AS2011}
which was used to create the atomic target, and the UCL-Belfast-Strathclyde version of the
R-matrix code\footnote{{See Badnell: R-matrix write-up on WWW. %
URL: \url{amdpp.phys.strath.ac.uk/UK_RmaX/codes/serial/WRITEUP}.}} \citep{BerringtonEN1995} to
carry out the scattering calculations.

The \AS\ code was used to generate the target radial functions which are used as an input to the
first stage of the R-matrix code. The radial data were produced using thirty-nine configurations
synthesised from seven orbitals; three spectroscopic (1s, 2s and 2p) and four correlation orbitals
($\po3s$, $\po3p$, $\po3d$ and $\po4d$). This configuration basis was previously used by
\citet{StoreySB2014} in their work on the collision strengths of [O\III] forbidden lines and is
given in Table~(1) of \citet{StoreySB2014}. The correlation orbitals are calculated in a Coulomb
potential with central charge 8$|\lambda_{nl}|$. An iterative optimisation variational procedure
was employed to obtain the orbital scaling parameters, $\lambda_{nl}$, as supplied in Table~(2) of
\citet{StoreySB2014}.

In the R-matrix scattering calculations, 72 target terms, listed in Table~(3) of
\citet{StoreySB2014}, were used. For the 36 energetically lowest states, experimental energies
obtained from \citet{MartinKM1993}, were used in the diagonal entries of the Hamiltonian in stage
STG3 of the R-matrix code, instead of the theoretically computed values obtained from stage STG1,
to provide accurate threshold energies and hence more accurate positions for resonances in the
photoionisation cross-sections.

\subsection{Energy Levels}\label{EnergyLevels}

The R-matrix calculation provides bound state energies for 674 levels with principal quantum number
$n\le10$ and orbital angular momentum quantum number $l\le4$. The Hamiltonian used in the
calculation of bound state energies includes electrostatic terms and one-body relativistic terms,
i.e. the mass and Darwin relativistic energy shifts and the spin-orbit interaction. Two-body spin
and orbit dependent terms are not included in this formulation. In Table~\ref{BoundTable1} we
compare the calculated ionisation energies with experiment for all terms up to the highest of the
2s$^2$2p$^2$3d electron configuration, 26 terms in all. The Table also shows the total
fine-structure splitting of each term compared to the experimental value.

For states of the principal series, \nlo2s2\nlo2p2($^3$P)$3l$ the agreement between theory and
experiment is excellent with a maximum difference of 0.9\% and differences that are generally much
less than 1\%. There are larger differences for the members of the ($^1$D)$nl$ and ($^1$S)$nl$
series that are in this energy range, reaching a maximum of 1.1\% and 2.8\% respectively. For low
$l$ we can expect $LS$ coupling to be a good approximation but to become increasingly inappropriate
as $l$ increases, independent of $n$.  \cite{LiuSBC1995} discuss the coupling schemes that are
appropriate for \op\ states and conclude that there is a partial breakdown of $LS$ coupling for
some of the ($^3$P)3d states and that $LS$ coupling is completely inappropriate for the ($^3$P)4f
levels. Departures from $LS$-coupling within a configuration ($^3$P)$nl$ depend, to a first
approximation, on the inverse of the energy separation between states of the same total angular
momentum $J$ belonging to different terms, and the magnitude of the spin-orbit parameter for the
orbital $nl$. The accuracy of the calculated spin-orbit parameter can be estimated by comparing the
total fine-structure splitting of individual terms with experiment. Table~\ref{BoundTable1} shows
that the calculated and experimental total fine-structure splittings of the ($^3$P)$3l$ terms agree
within 10\%.

The large differences between theory and experiment for the very small fine-structure splittings of
the \nlo2s2\nlo2p3 \SLP2Do and \SLP2Po terms arises because the spin-orbit parameter is zero for
the half-filled 2p shell and the observed level separations are due to two-body fine-structure
terms which are not included in our R-matrix formulation. In Table~\ref{BoundTable2} we compare
calculated and experimental energies for levels of the \nlo2s2\nlo2p2($^3$P)4f and 5f
configurations. These levels are no longer well represented in $LS$-coupling but appear as groups
attached to each of the \SL3P$_J$ parent levels. We show the statistically weighted mean energies
of each group and the total fine-structure splitting of the group.

\subsection{Bound-Bound and Bound-Free Radiative Data}

The R-matrix calculation provides electric dipole oscillator strengths between all 674 bound
states, which include spin-changing intercombination transitions. It also provides photoionisation
cross-sections resolved by final target level from which recombination coefficients can be
calculated. The energy mesh used to calculate the cross-sections is divided into two sections. From
the energy of the \nlo2s2\nlo2p2 \SLJ3P2 target state to just below the \nlo2s2\nlo2p2 \SLJ1D2
threshold, a mesh of variable step is used. The step length is varied to ensure that resonance
features in the cross-section, whose positions and widths have been previously determined, are well
resolved. Details of this approach can be found in \citet{KisieliusSDN1998}. This mesh terminates
at effective quantum number, $\nu=10$ relative to the \SLJ1D2 threshold and comprises 14244
energies. For higher energies the energy mesh between thresholds is defined by a fixed step,
$\Delta\nu=0.001$ in effective quantum number relative to the next threshold, giving rise to a
further 11569 energies. For $\nu \ge 10$ relative to any threshold, the cross-section is obtained
by Gailitis averaging with the mesh determined by $\Delta\nu=0.001$ relative to the next highest
threshold. The mesh extends to just above the 2s \nlo2p3 \SL3D$^{\rm o}$ target state to an energy
of 1.095~Rydberg. The calculations described by \cite{BastinThesis2006} and \cite{BastinS2006} were
also based on an R-matrix calculation of energy levels, oscillator strengths and photoionisation
cross-sections, albeit with a simpler scattering target than the one used here. However, a
re-examination of those photoionisation cross-sections shows that some of the cross-sections for
the more highly excited states display non-physical features, such as unexpected oscillations and
elevated continua. The results of the present calculation are therefore to be preferred.

For the 4f-3d and 3d-3p transition arrays we can compare our energies and branching ratios with
those of \citet{LiuSBC1995} who used the atomic structure code SUPERSTRUCTURE \citep{EissnerJN1974,
NussbaumerS1978} to calculate energy levels and dipole matrix elements for these transition arrays
in intermediate coupling. \citet{LiuSBC1995} made empirical corrections to the term and level
energies and spin-orbit parameters to bring the energies close to experiment. In the R-matrix
approach used here we can make corrections to the target energies but not to the final O$^+$ bound
state energies. In Table~\ref{LiuTable} we compare individual level energies with experiment and
with the results of \cite{LiuSBC1995} for the 3d and 4f configurations. The empirical adjustments
made by \citet{LiuSBC1995} mean that they obtained better fine-structure splittings than those
obtained here.  In Tables~\ref{LiuTable4a} and \ref{LiuTable4b} we compare the branching ratios for
selected stronger components of the 3d-3p and 4f-3d transition arrays computed by
\citet{LiuSBC1995} with the present work. For the larger branching ratios, greater than $0.1$,
there is general agreement between our results and those of \citet{LiuSBC1995} to within $3\%$ for
the 3d-3p transitions and to within $10\%$ for the 4f-3d transitions. However, \citet{LiuSBC1995}
point out that whereas the \SL4F, \SL2P and \SL2D terms of the \nlo2s2\nlo2p23d configuration are
well described by $LS$-coupling, the \SL4P, \SL4D and \SL2F terms are not. In
Table~\ref{LiuTable4b} we see that the largest differences between the present work and that of
\citet{LiuSBC1995} occurs for transitions involving the \SL4D and \SL2F terms and in particular the
\SLJ4D{5/2} and \SLJ2F{5/2} levels where the differences reach $40\%$. It is probable that the
branching ratios calculated by \citet{LiuSBC1995} with empirical corrections to level energies and
spin-orbit parameters are closer to reality than our {\it ab initio} results for these particular
transitions. We note, however, that the predicted strength of the corresponding recombination lines
also depends upon the relative populations of the O$^{2+}$ \SL3P$_J$ ground levels and these were
assumed to be in proportion to statistical weights by \citet{LiuSBC1995}. In the present
calculation the O$^{2+}$ \SL3P$_J$ level populations were explicitly included and accounted for in
the calculation of all the O$^+$ level populations. We return to this point below.

\subsection{Autoionisation Probabilities}

The calculation of the departure coefficients above the \SLJ3P0 ionisation limit and below the
\SLJ3P2 limit requires, in addition to the usual collisional-radiative processes, rates for
autoionisation and dielectronic capture. As shown in Figure~\ref{DRfig3}, these additional
processes apply for $n\ge63$ in the (\SLJ3P1)$nl$ series and $m\ge38$ in the (\SLJ3P2)$ml$ series.
However, in an R-matrix calculation there is no allowance for the competition between
autoionisation and radiative decay of the valence electron, so a simple integration of the
photoionisation cross-section does not correctly treat the dielectronic recombination {\it via}
these high $n,m$ states.  We therefore use AUTOSTRUCTURE \citep{EissnerJN1974, NussbaumerS1978,
AS2011} to calculate the autoionisation probabilities for selected $n,m <1000$ and $l\le30$ and to
avoid duplication of the dielectronic component due to the (\SLJ3P1)$nl$ and (\SLJ3P2)$ml$ Rydberg
states we begin the calculation of the R-matrix photoionisation cross-sections from the low-lying
states at the energy of the \otp\ \SLJ3P2 level. This provides the dielectronic contribution to the
recombination, allowing for all the usual collisional-radiative processes but not the non-resonant
or radiative recombination component, which we obtain by extrapolation of only the background
photoionisation cross-section from above the \SLJ3P2 threshold to the energy region between the
\SLJ3P0 and \SLJ3P2 states.

\section{Results and Discussion} \label{Data}

\subsection{Structure and Contents of Data File}

The O\II\ recombination lines and their theoretically-computed emission coefficients
(erg~cm$^3$~s$^{-1}$ per \otp\ ion per electron) up to $n=7$ are archived in a single file called
``OIIlines\_ABC'', where ABC refers to the three ``Cases'' for which emission coefficients were
calculated. Case A refers to the situation where all \op\ radiative transitions are assumed
optically thin. In Case B we assume that all radiative transitions terminating on the \op\
\SLPJ4So{3/2} ground level are sufficiently optically thick that the photons are re-absorbed
on-the-spot in the same transition. Hence, these decays are excluded from the calculation of level
populations. For Case C we also exclude decays to the \op\ \SLPJ2Do{3/2} and \SLPJ2Do{5/2}. This is
a simple and relatively crude treatment of the effects of optical depth in resonance transitions
but comparison with observations indicates that Case B is a much better approximation to reality in
the majority of PNe than Case A or Case C. We briefly discuss the possible relevance of Case C in
Section~\ref{ExampleResults}.

The ``OIIlines\_ABC'' file consists of two main sections preceded by a short header which consists
of a general description of the file structure and the abbreviations used to label and explain the
data blocks and their units. The structure of these sections is as follows:

\noindent 1. The first section consists of 8889 indexed text lines corresponding to the provided
8889 O\II\ transitions where each text line contains full identification of the corresponding
transition by its upper and lower level as well as its wavelength in \AA\ and other relevant atomic
designations.

\noindent 2. The second section consists of 8889 parts where each part corresponds to one
transition as indexed above. Each part is made of three blocks where each block corresponds to one
of the three Cases. Each one of these blocks is a rectangular array of emission coefficients of the
indexed transition arranged in 25 rows, as a function of $\log_{10}(T_e[{\rm K}])$ between 2.0--4.4
in steps of 0.1, and 16 columns as a function of $\log_{10}(N_e[{\rm cm}^{-3}])$ between 2.0--5.0
in steps of 0.2.

\noindent More details about the structure and contents of the data file are provided in the ReadMe
file associating the distributed data set.

The data set distributed as part of the current work also includes a file called ``OIImeta\_ABC''
containing the recombination coefficients (cm$^3$~s$^{-1}$ per \otp\ ion per electron) to the
ground and meta-stable levels of \op\ as functions of electron temperature and number density and
Case (A, B and C). The file contains 14 data blocks. The structure of the file is as follows:

\noindent 1. A block for the effective recombination coefficients of the ground level
\nlo2s2\nlo2p3 \SLPJ4So{3/2} in Case A.

\noindent 2. Four blocks for the effective recombination coefficients of the two levels
\nlo2s2\nlo2p3 \SLPJ2Do{5/2} and \SLPJ2Do{3/2} each in Case A and B.

\noindent 3. Six blocks for the effective recombination coefficients of the two levels
\nlo2s2\nlo2p3 \SLPJ2Po{3/2} and \SLPJ2Po{1/2} each in Case A, B and C.

\noindent 4. Three blocks for the total recombination coefficients, one block for each Case.

\noindent Each data block is structured as a function of electron temperature and number density in
$\log_{10}$ identical to the structure of the emission coefficient blocks in ``OIIlines\_ABC''
file, as described above. We note that direct recombination and all cascade contributions,
excluding those between the metastable states themselves, are included in the calculation of the
coefficients in the ``OIImeta\_ABC'' file.

\subsection{Data Mining}

The data in the ``OIIlines\_ABC'' file can be excavated using the supplied fortran 77 interactive
data server. This server reads the file and provides easy access to the lines and their emission
coefficients at the desired $T_e$ and $N_e$. When the required data are within the $T_e$ and $N_e$
ranges but are not at the grid points of $T_e$ and $N_e$ as described above, a 6-point Lagrange
interpolation in temperature and density is used to provide the desired data. Two main options are
offered by the data server: (1) Generate a line list, and (2) Extract emission coefficients for a
user supplied list of lines in user specified conditions.

As for the first option, the server offers two choices: (1a) List all the 8889 lines (between
367.97 and 3.90713$\times10^7$ \AA), and (1b) Specify a wavelength range for the list of lines
desired by the user. In each one of these cases (i.e. 1a and 1b) two files are generated that
contain the list of lines where the lines in one of these files (named ``OIIdata\_list\_eorder'')
are ordered in decreasing emission coefficient and in the other file (named
``OIIdata\_list\_worder'') in increasing order of wavelength; the full list in the latter file is
preceded by a list of only the strongest lines in the required wavelength range. The emission
coefficients in these files correspond to typical nebular conditions of $T_e=10^4$~K and
$N_e=10^4$~cm$^{-3}$.

Regarding the second option, the user is asked to supply a list of lines in a text file for which
the data are required. The user then has the choice between (2a) Providing a list of $T_e[{\rm
K}]$--$N_e[{\rm cm}^{-3}]$ pairs in a file for which the data are needed, or (2b) Providing a range
of $\log_{10}(T_e[{\rm K}])$ and $\log_{10}(N_e[{\rm cm}^{-3}])$ in a file defining a one- or
two-dimensional grid for which the data are required. In all of these options (i.e. 1a, 1b, 2a and
2b) the user has the choice to have the data for Case A, B or C. More details about the data server
and its input and output files are provided in the ReadMe file associating the distributed data and
code.

\section{Examples of Comparisons with Observations}\label{ExampleResults}

\subsection{Multiplet V1}
\begin{figure}
\centering{}
\includegraphics[scale=0.45]{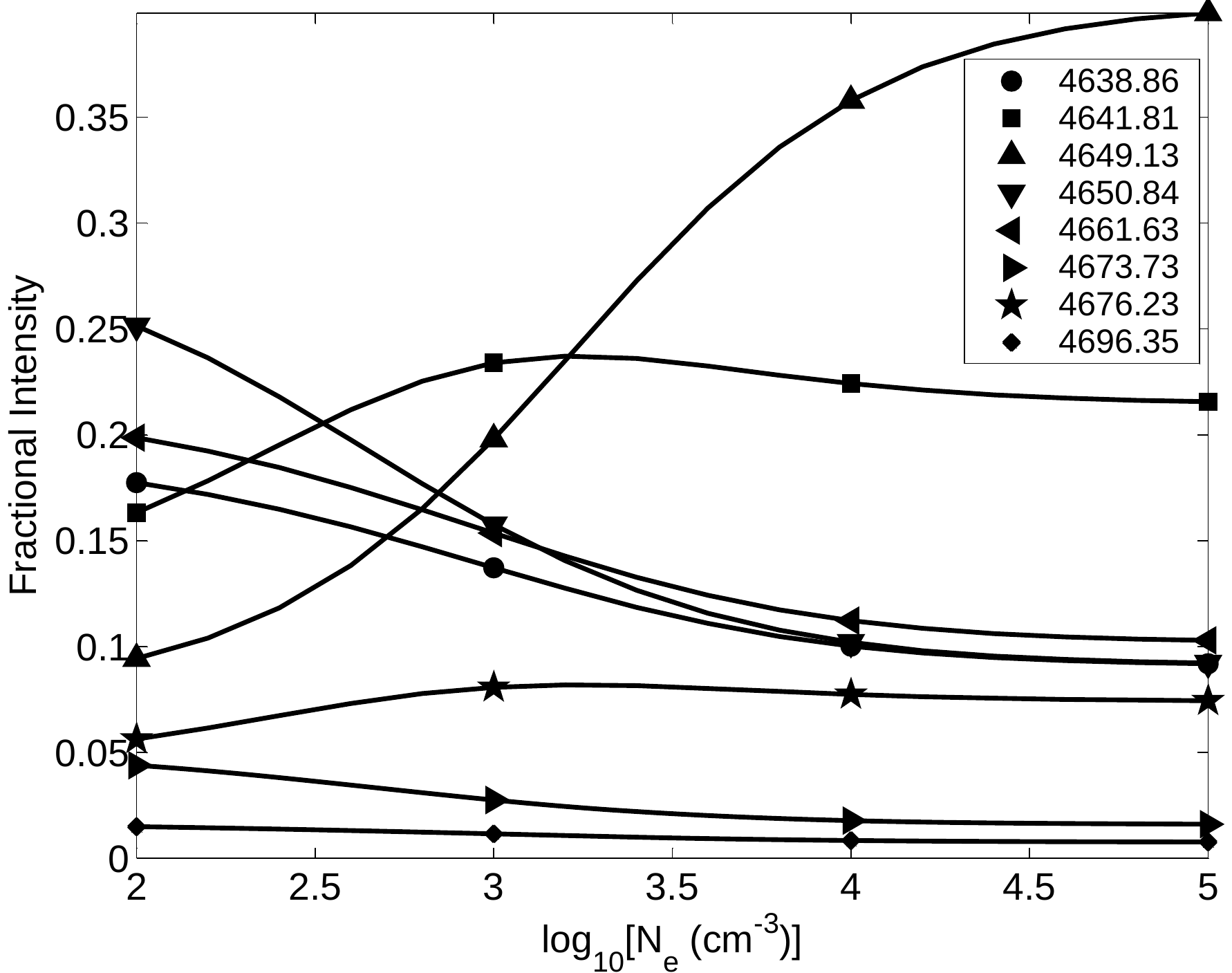}
\caption{Computed fractional intensities of the eight transitions of the V1 multiplet of  O\II\
at $T_e=10000$~K and in Case B as a function of $\log_{10}[N_e({\rm cm}^{-3})]$.} \label{V1TheoryT10000K}
\end{figure}

The lines of O\II\ multiplet V1 are among the brightest and most accessible spectroscopically. The
relative intensities of the components of V1 depend upon the electron density {\it via} the
distribution of population in the \SL3P$_J$ ground levels of \otp. This is illustrated in
Figure~\ref{V1TheoryT10000K} which shows the fractional intensities of the eight lines of V1 at a
temperature of 10000~K. The relative populations of the \SL3P$_J$ levels are approaching their
local thermodynamic equilibrium values at $N_e=10^5$~cm$^{-3}$. At the higher densities the
strongest component is the 4649.13~\AA\ line which, arising from the \SLPJ4Do{7/2} level, is
primarily formed by recombination from the \otp\ \SLJ3P2 level at this temperature. The relative
intensity of this component falls rapidly as the density and the population of the \SLJ3P2 level
fall. In their Figure~1, \cite{McnabbFLBS2013} also present the fractional intensities of the eight
lines of V1 over the same range of densities. Our results, which supersede those of
\cite{McnabbFLBS2013}, are nevertheless in good agreement for this case.

\begin{figure}
\centering{}
\includegraphics[scale=0.45]{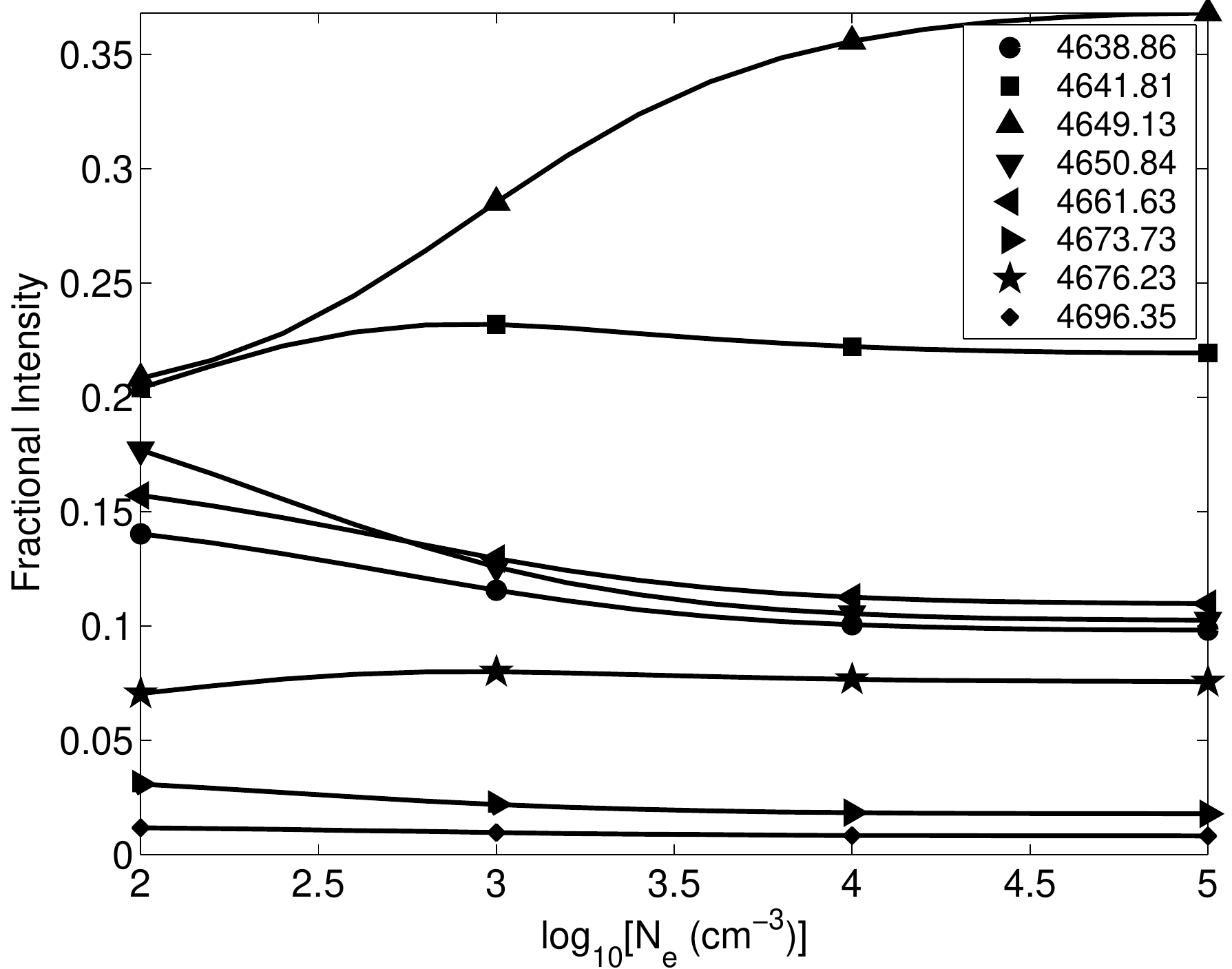}
\caption{Computed fractional intensities of the eight transitions of the V1 multiplet of the O\II\
at $T_e=1000$~K and in Case B as a function of $\log_{10}[N_e({\rm cm}^{-3})]$.} \label{V1TheoryT1000K}
\end{figure}

In Figure~\ref{V1TheoryT1000K} we show the fractional intensities for multiplet V1 at 1000~K. At
this temperature the density variation of 4649.13~\AA\ is much less marked than at 10000~K partly due to
the effect of dielectronic recombination through the high-$n$ Rydberg states converging on the
\SLJ3P1 and \SLJ3P2 levels. Through this mechanism ($^3$P$_2$)$nl$ levels can be populated even at
low electron density and, by radiative decays, eventually populate the upper level of the
4649.13~\AA\ line. This process is only significant when the mean thermal energy of the free
electrons is comparable to the \SL3P$_J$ level separations. Very similar results were shown by
\cite{McnabbFLBS2013} in their Figure~1.

\begin{figure}
\centering{}
\includegraphics[scale=0.45]{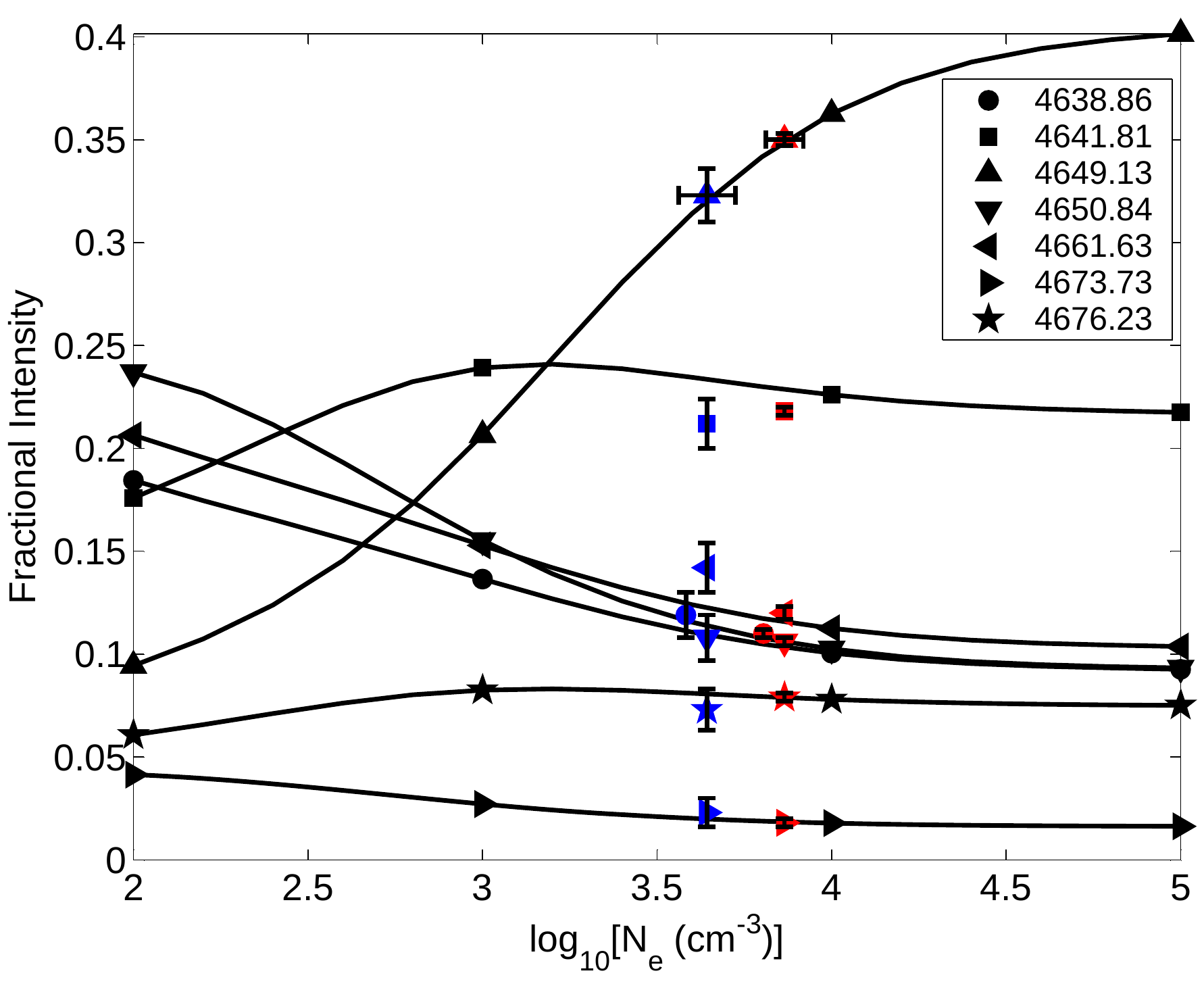}
\caption{Computed fractional intensities in Case B of seven transitions of the V1 multiplet of the
O\II\ recombination lines at $T_e=8320$~K as a function of $\log_{10}[N_e({\rm cm}^{-3})]$ with the
observational intensities of the H\II\ region Orion Nebula (M42) plotted at {\CR $N_e=7330^{+360}_{-350}$~cm$^{-3}$}
(red right) as obtained from \citet{BaldwinVVFMe2000} and at {\CR $N_e=4400^{+890}_{-740}$~cm$^{-3}$} (blue left) as
obtained from \citet{EstebanPGRPR2004}. Some points are displaced horizontally to avoid
overlapping.} \label{BaldwinEstebanV1}
\end{figure}

\citet{BaldwinVVFMe2000} and \citet{EstebanPGRPR2004} have published deep optical spectra of the
Orion Nebula, M42. If we assume an electron temperature we can use the measured and computed
relative intensities of the V1 lines to determine an electron density for the emitting regions. We
adopt the temperature of 8320~K derived by \citet{EstebanPGRPR2004} for the higher ionisation
species such as \otp\ in M42 and locate the electron density that gives the best fit (least
squares) between theory and observation for the seven lines of multiplet V1 for which measured
intensities are available. In Figure~\ref{BaldwinEstebanV1} we show the theoretical fractional
intensities plus the observed values from \citet{BaldwinVVFMe2000} and \citet{EstebanPGRPR2004}
plotted at the best fit electron densities. The vertical error bars are the error estimates given
by the authors and the horizontal error bar attached to the 4649.13~\AA\ line is the one sigma
uncertainty in the density derived from those observational errors. The agreement between theory
and observation at the best fit densities is excellent. The derived densities of 7330~cm$^{-3}$ and
4400~cm$^{-3}$ can be compared with the value of 8900~cm$^{-3}$ adopted for M42 by
\citet{EstebanPGRPR2004} from a weighted average of forbidden line density diagnostics.

\begin{figure}
\centering{}
\includegraphics[scale=0.45]{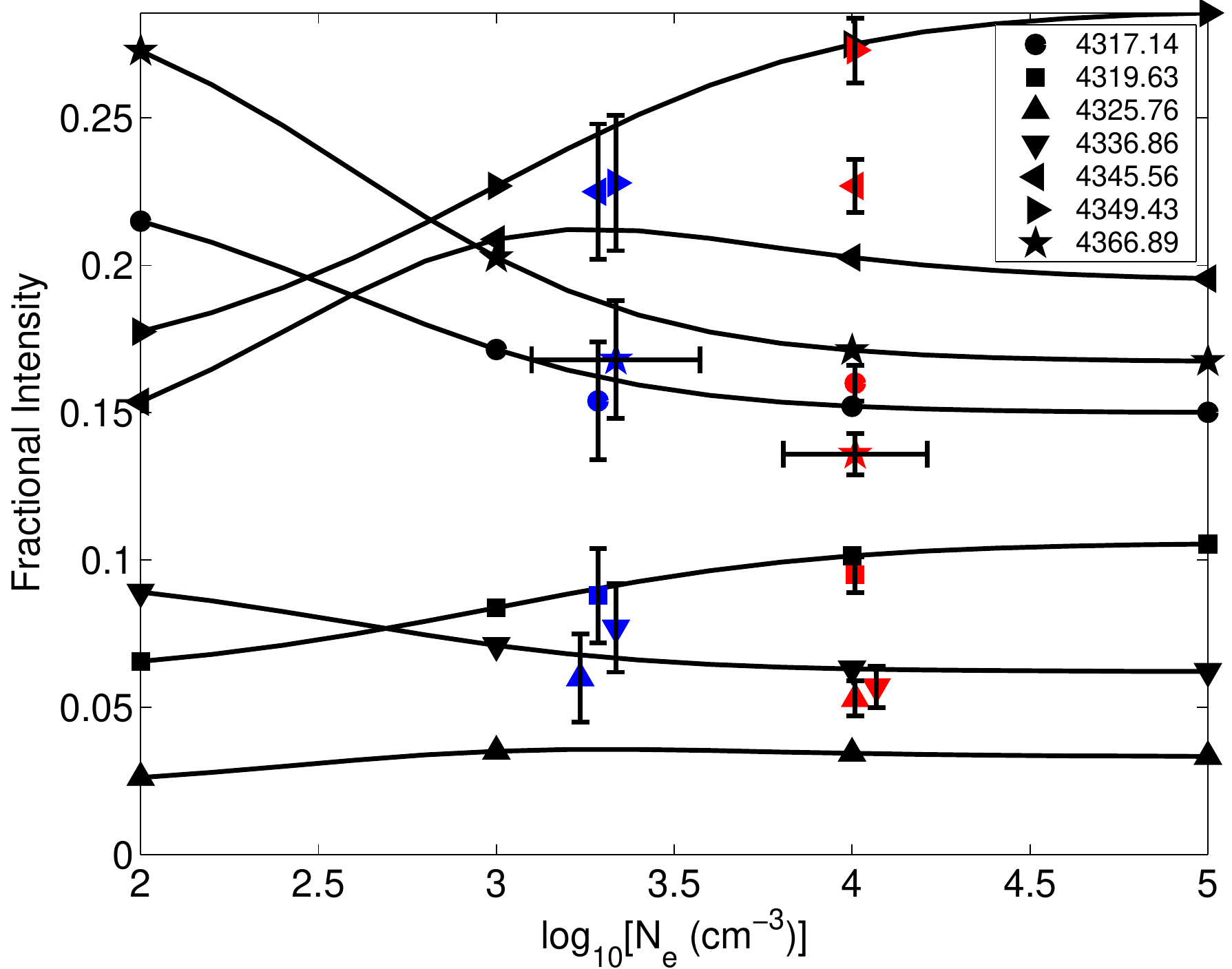}
\caption{Computed fractional intensities in Case B of seven transitions of the V2 multiplet of the
O\II\ recombination lines at $T_e=8320$~K as a function of $\log_{10}[N_e({\rm cm}^{-3})]$ with the
observational intensities of the H\II\ region Orion Nebula (M42) plotted at {\CR $N_e=10210^{+12910}_{-4040}$~cm$^{-3}$}
(red right) as obtained from \citet{BaldwinVVFMe2000} and at {\CR $N_e=2170^{+1640}_{-940}$~cm$^{-3}$} (blue left) as
obtained from \citet{EstebanPGRPR2004}. Some points are displaced horizontally to avoid
overlapping.} \label{BaldwinEstebanV2}
\end{figure}

In Figure~\ref{BaldwinEstebanV2} we again use the observations of \citet{BaldwinVVFMe2000} and
\citet{EstebanPGRPR2004} to derive electron densities which are the best fit to the seven observed
lines of multiplet V2. The lines of V2 are weaker than those of V1 and the observational errors are
consequently relatively larger, leading to larger uncertainties in the derived densities. Within
the error bars there is reasonable agreement between the results derived from multiplets V1 and V2.

\begin{figure}
\centering{}
\includegraphics[scale=0.45]{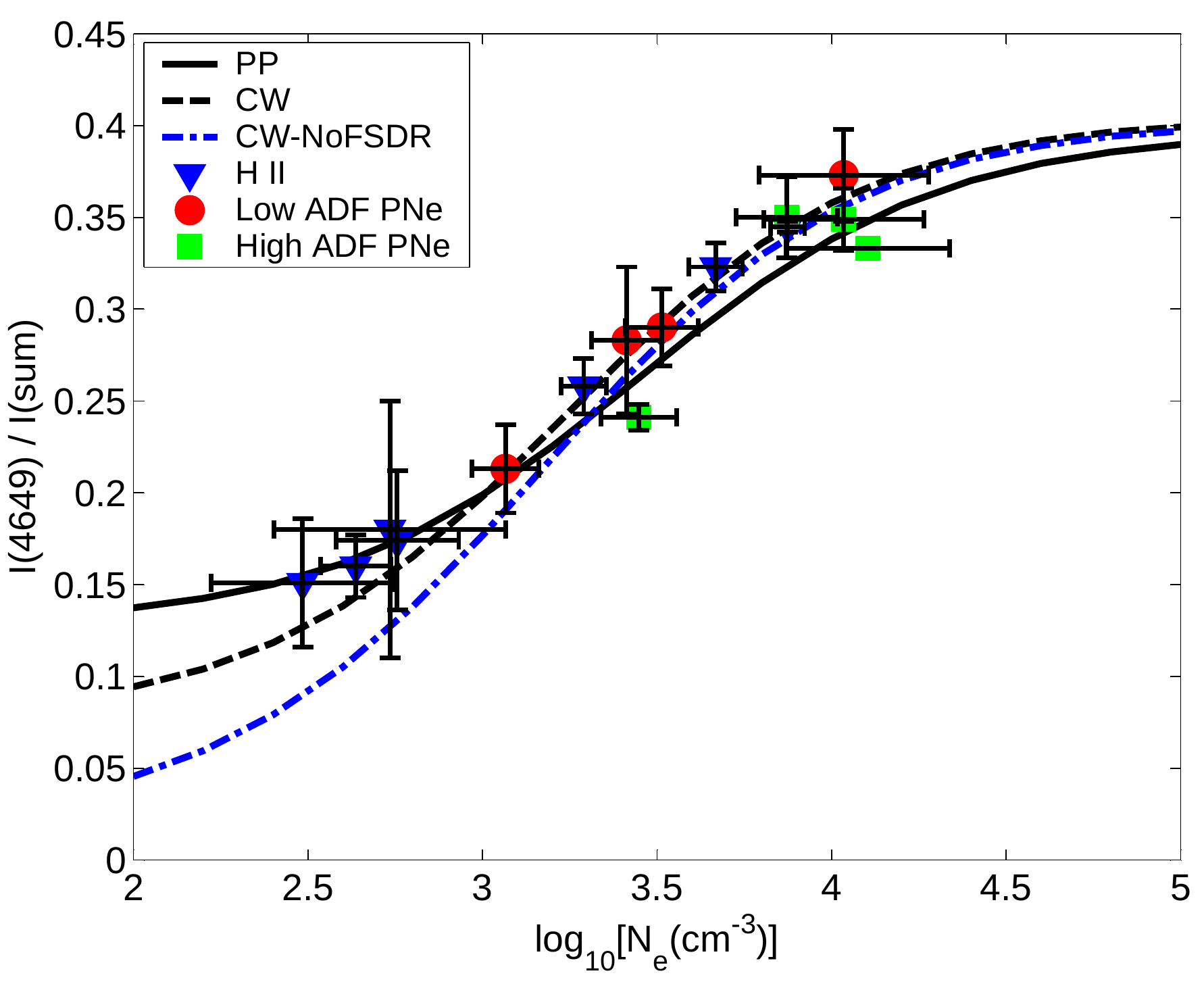}
\caption{The intensity of $\lambda4649.13$ relative to the total multiplet V1 intensity as a
function of $\log_{10}[N_e({\rm cm}^{-3})]$ from \citet{PeimbertP2005} (PP), the current work (CW)
and the current work neglecting fine-structure dielectronic recombination (CW-NoFSDR). The two
curves showing the present work correspond to $T_e=10000$~K in Case~B. See Table~\ref{tabObjects}
for information about the observational points.} \label{PeimbertP2005T10000K4d}
\end{figure}

In Figure~\ref{PeimbertP2005T10000K4d} we plot the intensity of V1 $\lambda4649.13$ relative to the
total V1 intensity as a function of electron number density. The results from the current work (CW)
were calculated in Case B at $10000$~K. We also show the effect on the ratio of switching off
fine-structure dielectronic recombination (CW-NoFSDR). The solid curve (PP) shows the empirical
results derived by \citet{PeimbertP2005} from the relationship between the observed values for a
range of objects and the density derived from forbidden line ratios, in practice mainly from
[Cl\III]. Figure~\ref{PeimbertP2005T10000K4d} also shows the results we obtain from the best fit of
the relative intensities of all the observed lines of multiplet V1 to the theoretical relative
intensities for three types of object: H~{\sc ii} regions, PNe with low ADF, which we define as ADF
$\le 2.5$, and PNe with high ADF $> 2.5$, where the ADF values were taken from
\citet{TsamisBLSD2004}, Table~9 (ORL/OPT for \otp). The densities resulting from the fits are given
in numerical form in Table~\ref{tabObjects}, {\CR which also lists the [O\III] optical forbidden
line temperature for each object}. The theoretical values used in the fitting procedure were all
calculated at $10000$~K, which may not be appropriate for the high ADF PNe {\CR where there is some
evidence that the recombination lines originate from a region at significantly lower temperature.
As discussed above and shown in Figure~\ref{V1TheoryT1000K} the variation of the $\lambda4649.13$
relative intensity with density is less pronounced at $1000$~K than at $10000$~K so that a given
observed ratio would generally imply a lower electron density than those obtained from the fits at
$10000$~K and listed in Table~\ref{tabObjects}.} We note that the empirical curve of
\citet{PeimbertP2005} is in good agreement with our theory except for low density objects, where
the [Cl\III] densities used by \cite{PeimbertP2005} tend to be lower than those derived here from
the recombination line theory.  There is therefore no clear indication from
Figure~\ref{PeimbertP2005T10000K4d} that the recombination lines are formed in a region of
significantly different density to the [Cl~{\sc iii}] forbidden lines, as might be expected if they
originated from different physical regions.

{\CR The density variations of the V1 lines and the lines of other O\II\ recombination multiplets
result primarily from the density variations of the populations of the \otp\ \SLJ3P{J} levels. The
critical density $N_i^{\rm crit}$ for a given level $i$, is given by:
\begin{equation}
N_i^{\rm crit} = \sum_{k<i} A_{ik}/\sum_{k\ne i}q_{ik}
\end{equation}
where $A_{ik}$ are radiative transition probabilities and $q_{ik}$ are rate coefficients for
collisional excitation or de-excitation. If the ambient density is much greater than $N_i^{\rm
  crit}$ for a given level, the level population will approach the value given by the Boltzmann
distribution. At $10^4$~K the critical densities of the \SLJ3P{1} and \SLJ3P{2} levels are
approximately 610 and 4350 cm$^{-3}$ respectively but for the level populations to approach within
10\% of Boltzmann values will require densities approximately 10 times the critical densities, i.e.
6100 and 43500 cm$^{-3}$. Typical PN (and H\II\ region) densities are lower than this so we can
conclude that the lines of all O\II\ recombination multiplets will normally exhibit a significant
dependence on density. Thus only in very high density nebulae can it be assumed that the \otp\
\SLJ3P{J} levels have a Boltzmann distribution of population and that the previous $LS$-coupling
theoretical results (e.g. \cite{Storey1994}) can be used to infer abundances from individual lines
without knowledge of the electron density.

An additional consideration is that critical densities decrease with decreasing temperature, so
that at $10^3$~K for example, the values of $10\times N_i^{\rm crit}$ for the \SLJ3P{1} and
\SLJ3P{2} levels of \otp\ are 2500 and 15500 cm$^{-3}$ respectively. These values are significantly
lower than those at $10^4$~K but not sufficiently low that a Boltzmann distribution can be safely
assumed.

It is useful to apply the same argument to the recombination lines of N\II\ and Ne\II, where we
find that $10\times N_i^{\rm crit}$ is 20500~cm$^{-3}$ for the \SLPJ2Po{3/2} level of \ntp\ and
$1.96\times10^6$ and $2.81\times10^5$~cm$^{-3}$ for the \SLJ3P{1} and \SLJ3P{0} levels of \netp.
Therefore we anticipate that all recombination multiplets of N\II\ and Ne\II\ will show significant
density dependence at nebular densities.}

\subsection{Density and Temperature Determinations}

\begin{figure*}
\centering{}
\includegraphics[scale=0.75]{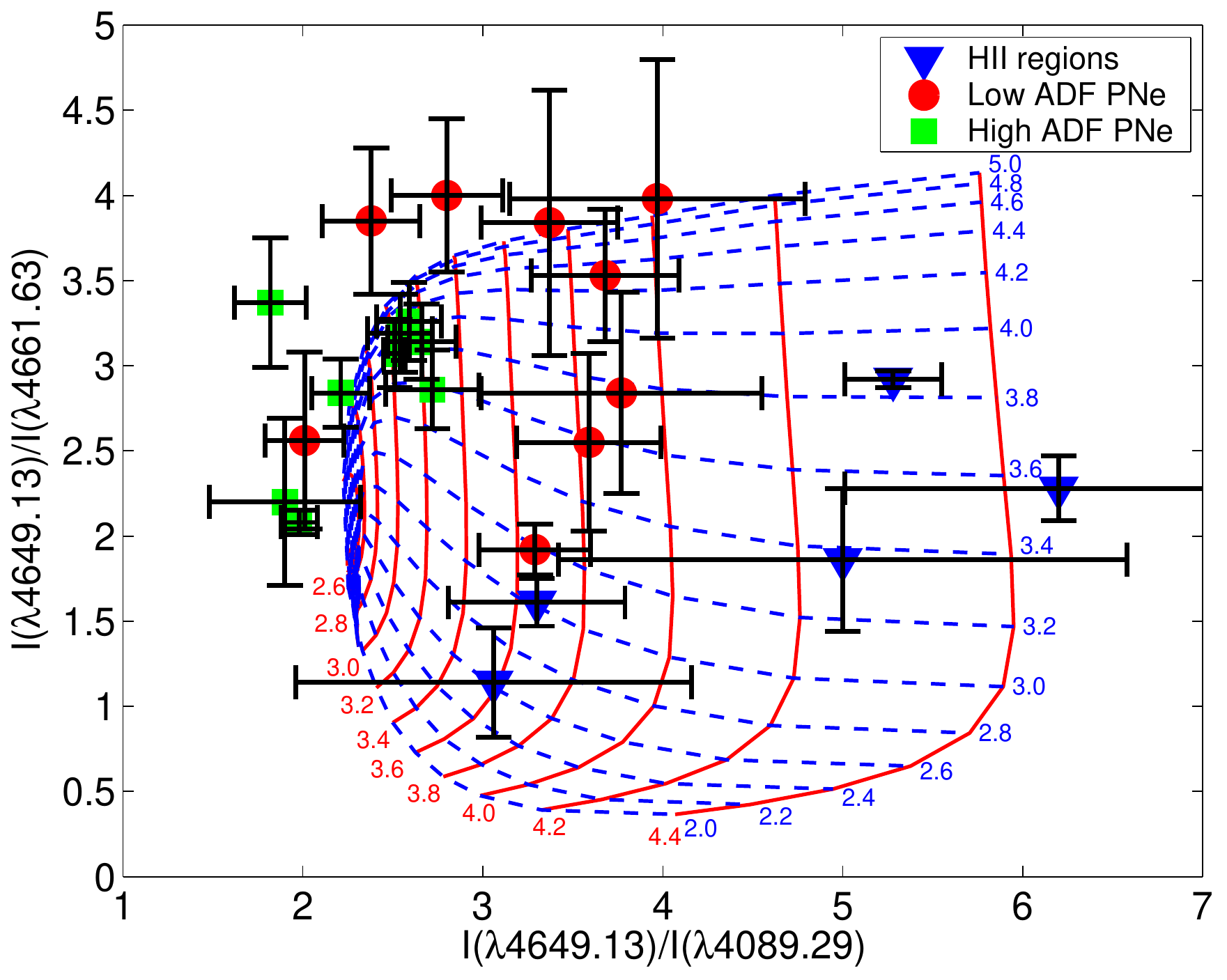}
\caption{The ratio of our theoretical line intensities $I(\lambda4649.13)/I(\lambda4661.63)$ versus
the ratio of our theoretical line intensities $I(\lambda4649.13)/I(\lambda4089.29)$ in Case~B and
as functions of $\log_{10}[T_e({\rm K})]$ (solid red) and $\log_{10}[N_e({\rm cm}^{-3})]$ (dashed
blue). See the text for the sources of the observational points.} \label{McNabbFLBS2013Plot}
\end{figure*}

In the previous section we chose to use observations of M42 to initially illustrate the
determination of electron density because there is broad agreement about the electron temperatures
in M42 and H\II\ regions generally. The same cannot be said about PNe where some models propose
that recombination lines are formed in regions of much lower temperature than those derived from
forbidden line diagnostics. It is therefore of interest to seek diagnostics that determine both the
density and the temperature of the regions where recombination lines are formed. In
Figure~\ref{McNabbFLBS2013Plot} we plot a predominantly density sensitive line ratio
$\lambda$4649.13/$\lambda$4661.63 against one which is mainly sensitive to temperature
$\lambda$4649.13/$\lambda$4089.29, in Case B. The former pair of lines are from the V1 multiplet
and the latter pair comprise the ratio of a V1 line with the strongest line from the 4f-3d
transition array, $\lambda$4089.29. In the latter case, both lines arise from the highest $J$ of
the transition array and hence depend strongly on the population of the $^3$P$_2$ ground level of
\otp, so that there is relatively little sensitivity to density in the ratio. Figure~2 of
\cite{McnabbFLBS2013} plots the same pair of ratios based on the earlier calculations of
\cite{BastinThesis2006}, which shows some minor differences to our Figure~\ref{McNabbFLBS2013Plot},
particularly in the values of the temperature sensitive $\lambda$4649.13/$\lambda$4089.29 ratio. As
stated above, the earlier results contained defective photoionisation cross-sections for some
states and are therefore superseded by the present work.

{\CR \cite{PeimbertP2013} have suggested that, in low resolution spectra, the intensity of the
$\lambda$4089.29 line may be affected by blending with a Si~{\sc iv} line at 4088.86\AA\ which,
once corrected for, would potentially increase the observed $\lambda$4649.13/$\lambda$4089.29 ratio
and result in deriving higher electron temperatures from this ratio. The Si~{\sc iv}
$\lambda$4088.86 line is part of a doublet with the other, weaker component at $\lambda$4116.10,
and a line at this wavelength has indeed been observed in M42 and several PNe
\citep{PeimbertP2013}, although other identifications have been proposed. Thus caution should be
exercised in deriving electron temperatures using a single ratio such as
$\lambda$4649.13/$\lambda$4089.29.}

{\CR Notwithstanding the possible blending issues, we show observational} values for 23 objects
including H\II\ regions, low ADF PNe and high ADF PNe, defined as in the previous section. The
sources of the observational data are as follows.  H\II\ regions: \cite{BaldwinVVFMe2000},
\cite{TsamisBLDS2003}, \cite{EstebanPGRPR2004}, \cite{RojasEPRRP2004}. Low ADF PNe:
\cite{HyungAFLK2000}, \cite{TsamisBLDS2003b}, \cite{RojasMLSMDT2015}. High ADF PNe:
\cite{LiuSBC1995}, \cite{LiuSBDCB2000}, \cite{LiuLBDS2001}, \cite{TsamisBLDS2003b},
\cite{LiuBZBS2006}, \cite{FangL2011}, \cite{McnabbFL2016}. The $\lambda4089.29$ line which is used
to constrain the temperature is a 4f-3d transition and is relatively weak. This is reflected in the
large error bars on the $\lambda4649.13/\lambda4089.29$ ratio particularly for the H\II\ regions
while in the high ADF PNe, which exhibit stronger recombination lines, the error bars are
relatively smaller. The three classes of objects broadly occupy different regions of the diagram.
Most marked is the tendency for the high ADF PNe to cluster in the very low temperature region of
the plot. The low ADF PNe, on the other hand, mainly occupy the region where temperatures are
closer to 10000~K, although there are exceptions. The data for H\II\ regions show a surprisingly
large scatter in terms of temperature, which may be partially attributable to the intrinsic
weakness of recombination lines when ADF factors are close to unity.  We emphasise also that the
theoretical results in Figure~\ref{McNabbFLBS2013Plot} were calculated in Case B which may not be
the best approximation for H\II\ regions which have large spatial extent and low ionisation. This
may lead to sufficient population to build up in the \SLPJ2Do{3/2} and \SLPJ2Do{5/2} levels of \op\
for radiative transitions to these two levels to become optically thick and hence for Case~C to be
more appropriate than Case~B.

\section{Conclusions} \label{Conclusions}

We have calculated emission coefficients for recombination lines of O\II\ over a range of electron
temperatures and densities in Cases A, B and C, suitable for application to the spectra of
photoionised nebulae. The calculation treats the important states in intermediate coupling and uses
rates for bound-bound and bound-free radiative processes computed with the R-matrix method of
electron scattering. The populations of the fine-structure levels of the ground term of \otp\ are
explicitly included, as is the effect of dielectronic recombination {\it via} high-$n$ Rydberg
states converging on these levels. To the best of our knowledge, these processes have not been
included in any previous calculation of recombination coefficients for O\II. Based on the above
calculations, a large data set that comprises 8889 recombination lines with their emission
coefficients as functions of electron temperature, number density and Case (A, B and C) accompanies
this paper. An interactive data server, in the form of a fortran 77 code, is provided for the
convenience of the users to explore the data and obtain Lagrange interpolated values between the
explicitly provided ones. We also provide a data set of coefficients for recombination to the
ground level and the four metastable levels of \op\ plus the total \otp + e$^-$ recombination
coefficients as functions of electron temperature and density and Case, for use in modelling the
oxygen ionisation balance and calculating the recombination contribution to the excitation of the
O\II\ forbidden lines.

Comparison with observations showed that the lines of multiplet V1, the brightest of the O\II\
recombination spectrum, provide a novel means of determining electron density, at least in those
objects where the electron temperature is reasonably well known. We have also shown that there is
good agreement between our calculated relative strengths of multiplet V1 as a function of electron
density and empirical results derived from observations and CEL densities. {\CR The density
dependence of the lines of the V1 multiplet at typical nebular densities is a result of these
densities being below the critical densities for the ground levels of \otp, and similar density
dependence is therefore expected to appear in the components of all recombination multiplets of
O\II\, and in all ions where a similar situation prevails.} Determining electron temperature from
O\II\ recombination lines is more problematic due to the relative weakness of the 4f-3d lines
which, when compared to multiplet V1, show sensitivity to temperature. There does, however, seem to
be evidence that, for the group of high ADF PNe considered here, the recombination lines are formed
at a markedly lower temperature than is derived from CELs.

\section{Acknowledgment}

{\CR The authors would like to thank the reviewer Professor M. Peimbert for comments
and corrections and for pointing out the possible blend of
the $\lambda$4089.29 line with a Si~{\sc iv} line.} The work of PJS was supported in part by the
STFC (grant ST/J000892/1).

\section{Statement}

The complete data generated in this work can be obtained in electronic format with full precision
from the Centre de Donn\'{e}es astronomiques de Strasbourg (CDS) database. The fortran 77 data
server is compiled and tested thoroughly using gfortran, f77, and Intel fortran compilers on Ubuntu
12.04 and Scientific Linux platforms. Representative sample results from all these compilers and on
all those platforms are compared and found to be identical within the stated numerical accuracy.

\onecolumn

\clearpage


\begin{table} 
\caption{The 26 lowest bound terms of the O$^+$ ion, which include all levels up to the highest of
the \nlo1s2\nlo2s2\nlo2p2(\SL3P)3d configuration in experimental energy order, and their
experimental ($E_{\rm{ex}}$) and theoretical ($E{_{\rm{th}}}$) energies in Rydberg relative to the
first ionisation limit (\nlo1s2\nlo2s2\nlo2p2 \SLPJ3P{}0), as well as the total fine-structure
splitting in cm$^{-1}$ of each term experimentally (TSTE) and theoretically (TSTT). The column
\%$E$ represents the percentage relative difference between the experimental and theoretical
energies $\left[\frac{100(E_{\rm{ex}}-E_{\rm{th}})}{E_{\rm{ex}}}\right]$ while the \%S column
represents a similar percentage for the splitting. The experimental energies are obtained from the
NIST database (\url{www.nist.gov}) while the theoretical energies are obtained from the bound stage
of the R-matrix code. The 1s$^2$ core is suppressed from all configurations. \label{BoundTable1}}
\vspace{-0.2cm}
\begin{center}
\begin{tabular}{clcrrrrr}
\hline
{\bf Index} & {\bf Term} & {\boldmath $E_{\rm{ex}}$} & {\boldmath $E_{\rm{th}}$} &  {\boldmath \%$E$} & {\bf TSTE} & {\bf TSTT} &  {\bf \%S} \\
\hline
1 & \nlo2s2 \nlo2p3 \SLP4So &  -2.581409 &  -2.587200 &      -0.22 &   0.0 &   0.0 &        0.0 \\

2 & \nlo2s2 \nlo2p3 \SLP2Do &  -2.337020 &  -2.333770 &       0.14 &   20.0  &  -8.8 &      143.9 \\

3 & \nlo2s2 \nlo2p3 \SLP2Po &  -2.212625 &  -2.203115 &       0.43 &   2.0  &  -6.6 &      430.9 \\

4 & 2s \nlo2p4 \SLP4P{} &  -1.488604 &  -1.487096 &       0.10 &   245.6  &  228.2 &        7.1 \\

5 & 2s \nlo2p4 \SLP2D{} &  -1.068751 &  -1.052943 &       1.48 &   8.0  &  -6.6 &      181.9 \\

6 & \nlo2s2 \nlo2p2(\SL3P)3s \SLP4P{} &  -0.891730 &  -0.893485 &      -0.20 &   263.8  &  276.0 &       -4.6 \\

7 & \nlo2s2 \nlo2p2(\SL3P)3s \SLP2P{} &  -0.858863 &  -0.858604 &       0.03 &   180.0  &  191.2 &       -6.2 \\

8 & 2s \nlo2p4 \SLP2S{} &  -0.797926 &  -0.769659 &       3.54 &   0.0  &  0.0 &        0.0 \\

9 & \nlo2s2 \nlo2p2(\SL3P)3p \SLP2So &  -0.722911 &  -0.725836 &      -0.40 &   0.0  &  0.0 &        0.0 \\

10 & \nlo2s2 \nlo2p2(\SL3P)3p \SLP4Do &  -0.695903 &  -0.698399 &      -0.36 &   271.7  &  287.8 &       -5.9 \\

11 & \nlo2s2 \nlo2p2(\SL1D)3s \SLP2D{} &  -0.695301 &  -0.690584 &       0.68 &   1.0  &  -0.2 &      121.1 \\

12 & \nlo2s2 \nlo2p2(\SL3P)3p \SLP4Po &  -0.681940 &  -0.684113 &      -0.32 &   138.1  &  148.7 &       -7.7 \\

13 & \nlo2s2 \nlo2p2(\SL3P)3p \SLP2Do &  -0.652751 &  -0.652963 &      -0.03 &   190.6  &  196.3 &       -3.0 \\

14 & \nlo2s2 \nlo2p2(\SL3P)3p \SLP4So &  -0.648007 &  -0.649264 &      -0.19 &   0.0  &  0.0 &        0.0 \\

15 & 2s \nlo2p4 \SLP2P{} &  -0.643687 &  -0.621974 &       3.37 &   168.4  &  162.1 &        3.8 \\

16 & \nlo2s2 \nlo2p2(\SL3P)3p \SLP2Po &  -0.629299 &  -0.628628 &       0.11 &   59.8  &  65.0 &       -8.7 \\

17 & \nlo2s2 \nlo2p2(\SL1D)3p \SLP2Fo &  -0.496954 &  -0.492036 &       0.99 &   23.6  &  23.4 &        1.0 \\

18 & \nlo2s2 \nlo2p2(\SL1D)3p \SLP2Do &  -0.485860 &  -0.481127 &       0.97 &   21.4  &  15.5 &       27.6 \\

19 & \nlo2s2 \nlo2p2(\SL1S)3s \SLP2S{} &  -0.479897 &  -0.466390 &       2.81 &   0.0  &  0.0 &        0.0 \\

20 & \nlo2s2 \nlo2p2(\SL3P)3d \SLP4F{} &  -0.472386 &  -0.471490 &       0.19 &   234.1  &  238.9 &       -2.0 \\

21 & \nlo2s2 \nlo2p2(\SL3P)3d \SLP4P{} &  -0.462644 &  -0.461378 &       0.27 &   139.8  &  135.0 &        3.4 \\

22 & \nlo2s2 \nlo2p2(\SL1D)3p \SLP2Po &  -0.462528 &  -0.457352 &       1.12 &   46.6  &  45.8 &        1.9 \\

23 & \nlo2s2 \nlo2p2(\SL3P)3d \SLP4D{} &  -0.460410 &  -0.459278 &       0.25 &   42.2  &  38.7 &        8.2 \\

24 & \nlo2s2 \nlo2p2(\SL3P)3d \SLP2F{} &  -0.459102 &  -0.458015 &       0.24 &   162.9  &  160.9 &        1.3 \\

25 & \nlo2s2 \nlo2p2(\SL3P)3d \SLP2P{} &  -0.453929 &  -0.449836 &       0.90 &   114.1  &  112.9 &        1.0 \\

26 & \nlo2s2 \nlo2p2(\SL3P)3d \SLP2D{} &  -0.445033 &  -0.443454 &       0.35 &   51.8  &  48.6 &        6.2 \\
\hline
\end{tabular}
\end{center}
\end{table}


\begin{table} 
\caption{The mean energies and total fine-structure splittings of the levels of the configurations
\nlo1s2\nlo2s2\nlo2p2(\SL3P$_J$)4f, 5f relative to the first ionisation limit. TSTE and TSTT are
the experimental and theoretical total fine-structure splitting of each group of levels while \%S
is the percentage difference. Other aspects of the table are as explained in Table
\ref{BoundTable1}. \label{BoundTable2}}
\begin{center}
\begin{tabular}{lrrrrrrr}
\hline
{\bf Configuration} & {\boldmath $E_{\rm{ex}}$} & {\boldmath $E_{\rm{th}}$} &  {\boldmath \%$E$} & {\bf TSTE} & {\bf TSTT} &  {\bf \%S} \\
\hline
\nlo2s2 \nlo2p2(\SLJ3P0)4f & -0.251329 & -0.251097 & 0.09 & 1.4  & 1.9 &  36.0 \\

\nlo2s2 \nlo2p2(\SLJ3P1)4f & -0.250321 & -0.250093 & 0.09 & 73.5  & 70.1 & -4.6   \\

\nlo2s2 \nlo2p2(\SLJ3P2)4f & -0.247971 & -0.247825 & 0.06 & 230.9  & 210.5 &  -8.8 \\

\nlo2s2 \nlo2p2(\SLJ3P0)5f & -0.160683 & -0.160573 & 0.07 & 1.5  & 1.9 &  27.0 \\

\nlo2s2 \nlo2p2(\SLJ3P1)5f & -0.159670 & -0.159557 & 0.06 & 49.1  & 46.6 & -5.1 \\

\nlo2s2 \nlo2p2(\SLJ3P2)5f & -0.157735 & -0.157653 & 0.05 & 111.8  & 100.0 &  -10.6 \\

\hline
\end{tabular}
\end{center}
\end{table}


\begin{table} 
\caption{Comparison of level energies in cm$^{-1}$ within the \nlo1s2\nlo2s2\nlo2p2(\SL3P)4f and
\nlo1s2\nlo2s2\nlo2p2(\SL3P)3d configurations between the experimental values ($E_{\rm ex}$), the
work of \citet{LiuSBC1995} (LSBC), and the current work (CW), where the given energies are relative
to the lowest level in each configuration. The first three columns are reproduced from Table (3) of
\citet{LiuSBC1995}. \label{LiuTable}} \vspace{-0.2cm}
\begin{center}
\begin{tabular}{lrrr}
\hline
{\bf Level} & {\boldmath $E_{\rm ex}$} & {\bf LSBC} &   {\bf CW} \\
\hline
4f D[3]${\rm ^o}_{5/2}$ &        0.0 &        0.0 &        0.0 \\

4f D[3]${\rm ^o}_{7/2}$ &        1.4 &        1.0 &        1.9 \\

4f G[3]${\rm ^o}_{5/2}$ &       66.2 &       66.3 &       67.4 \\

4f G[3]${\rm ^o}_{7/2}$ &       69.4 &       69.5 &       71.0 \\

4f D[2]${\rm ^o}_{3/2}$ &      122.7 &      122.7 &      124.2 \\

4f D[2]${\rm ^o}_{5/2}$ &      123.5 &      123.2 &      125.4 \\

4f G[4]${\rm ^o}_{9/2}$ &      137.7 &      137.7 &      135.2 \\

4f G[4]${\rm ^o}_{7/2}$ &      139.6 &      139.8 &      137.5 \\

4f D[1]${\rm ^o}_{3/2}$ &      222.3 &      221.9 &      227.3 \\

4f D[1]${\rm ^o}_{1/2}$ &      222.4 &      222.4 &      226.9 \\

4f G[5]${\rm ^o}_{11/2}$ &      287.5 &      287.5 &      286.8 \\

4f G[5]${\rm ^o}_{9/2}$ &      293.6 &      294.2 &      293.7 \\

4f F[2]${\rm ^o}_{3/2}$ &      393.7 &      393.3 &      378.0 \\

4f F[2]${\rm ^o}_{5/2}$ &      397.8 &      397.3 &      382.8 \\

4f F[3]${\rm ^o}_{7/2}$ &      433.3 &      432.9 &      417.5 \\

4f F[3]${\rm ^o}_{5/2}$ &      435.8 &      435.4 &      417.5 \\

4f F[4]${\rm ^o}_{9/2}$ &      446.1 &      445.8 &      429.5 \\

4f F[4]${\rm ^o}_{7/2}$ &      453.3 &      452.8 &      437.7 \\

           &            &            &            \\

3d \SLPJ4F{}{3/2}   &        0.0 &        0.0 &        0.0 \\

3d \SLPJ4F{}{5/2}   &       53.9 &       54.5 &       55.3 \\

3d \SLPJ4F{}{7/2}   &      131.8 &      133.0 &      134.8 \\

3d \SLPJ4F{}{9/2}   &      234.1 &      236.1 &      238.9 \\

3d \SLPJ4P{}{5/2}   &     1166.6 &     1158.6 &     1211.5 \\

3d \SLPJ4P{}{3/2}   &     1239.8 &     1234.2 &     1282.3 \\

3d \SLPJ4P{}{1/2}   &     1306.4 &     1303.5 &     1346.4 \\

3d \SLPJ4D{}{1/2}  &     1415.5 &     1415.2 &     1447.6 \\

3d \SLPJ4D{}{3/2}   &     1449.9 &     1449.6 &     1481.0 \\

3d \SLPJ4D{}{5/2}  &     1451.4 &     1453.0 &     1478.7 \\

3d \SLPJ4D{}{7/2}   &     1457.7 &     1460.9 &     1486.4 \\

3d \SLPJ2F{}{5/2}   &     1500.2 &     1502.0 &     1525.3 \\

3d \SLPJ2F{}{7/2}   &     1663.1 &     1665.6 &     1686.2 \\

3d \SLPJ2P{}{3/2}   &     2134.4 &     2131.4 &     2488.5 \\

3d \SLPJ2P{}{1/2}   &     2248.4 &     2260.5 &     2601.4 \\

3d \SLPJ2D{}{3/2}   &     3106.7 &     3105.4 &     3186.6 \\

3d \SLPJ2D{}{5/2}   &     3158.5 &     3161.7 &     3235.2 \\
\hline
\end{tabular}
\end{center}
\end{table}


\begin{table} 
\caption{Branching ratios for the strongest 3d-3p transitions in Case B as obtained by
\citet{LiuSBC1995} (LSBC) and the current work (CW) where $\lambda$ is the air wavelength.
\label{LiuTable4a}}
\begin{center}
\begin{tabular}{cccccc}
\hline
                         \multicolumn{3}{c}{{\bf Transition}} & {\boldmath $\lambda$[\AA]} & {\bf LSBC} &   {\bf CW} \\
\hline
\SLJ4F{9/2} &          - &       \SLPJ4Do{7/2} &    4075.86 &      1.000 &      1.000 \\

\SLJ4F{7/2} &          - &       \SLPJ4Do{7/2} &    4092.93 &      0.119 &      0.119 \\

\SLJ4F{7/2} &          - &       \SLPJ4Do{5/2} &    4072.15 &      0.864 &      0.862 \\

\SLJ4F{5/2} &          - &       \SLPJ4Do{7/2} &    4106.02 &      0.007 &      0.007 \\

\SLJ4F{5/2} &          - &       \SLPJ4Do{5/2} &    4085.12 &      0.216 &      0.216 \\

\SLJ4F{5/2} &          - &       \SLPJ4Do{3/2} &    4069.89 &      0.761 &      0.759 \\

\SLJ4F{3/2} &          - &       \SLPJ4Do{5/2} &    4094.14 &      0.016 &      0.017 \\

\SLJ4F{3/2} &          - &       \SLPJ4Do{3/2} &    4078.84 &      0.264 &      0.264 \\

\SLJ4F{3/2} &          - &       \SLPJ4Do{1/2} &    4069.62 &      0.715 &      0.715 \\

\SLJ4D{7/2} &          - &       \SLPJ4Do{7/2} &    3882.19 &      0.137 &      0.134 \\

\SLJ4D{7/2} &          - &       \SLPJ4Do{5/2} &    3863.50 &      0.016 &      0.016 \\

\SLJ4D{7/2} &          - &       \SLPJ2Do{5/2} &    4751.28 &      0.021 &      0.021 \\

\SLJ4D{7/2} &          - &       \SLPJ4Po{5/2} &    4119.22 &      0.378 &      0.370 \\

\SLJ4P{5/2} &          - &       \SLPJ4Do{5/2} &    3907.46 &      0.054 &      0.059 \\

\SLJ4P{5/2} &          - &       \SLPJ4Po{5/2} &    4169.23 &      0.153 &      0.133 \\

\SLJ4P{5/2} &          - &       \SLPJ4Do{3/2} &    3893.52 &      0.013 &      0.014 \\

\SLJ4P{5/2} &          - &       \SLPJ4Po{3/2} &    4153.30 &      0.449 &      0.458 \\

\SLJ4P{5/2} &          - &       \SLPJ4So{3/2} &    4924.53 &      0.289 &      0.283 \\

\hline
\end{tabular}
\end{center}
\end{table}


\begin{table} 
\caption{Branching ratios for the strongest 4f-3d transitions. All other aspects are as for Table
\ref{LiuTable4a}. \label{LiuTable4b}}
\begin{center}
\begin{tabular}{cccccc}
\hline
            \multicolumn{ 3}{c}{{\bf Transition}} & {\boldmath $\lambda$[\AA]} & {\bf LSBC} &   {\bf CW} \\
\hline
\hspace{0.13cm}G[5]$^{\rm o}_{11/2}$ &          - &       \SLJ4F{9/2} &    4089.29 &      1.000 &      0.999 \\

G[5]$^{\rm o}_{9/2}$ &          - &       \SLJ4F{9/2} &    4088.27 &      0.023 &      0.023 \\

G[5]$^{\rm o}_{9/2}$ &          - &       \SLJ4F{7/2} &    4071.24 &      0.204 &      0.204 \\

G[5]$^{\rm o}_{9/2}$ &          - &       \SLJ2F{7/2} &    4342.01 &      0.702 &      0.695 \\

G[5]$^{\rm o}_{9/2}$ &          - &       \SLJ4D{7/2} &    4303.61 &      0.072 &      0.076 \\

G[4]$^{\rm o}_{9/2}$ &          - &       \SLJ4F{9/2} &    4114.51 &      0.019 &      0.017 \\

G[4]$^{\rm o}_{9/2}$ &          - &       \SLJ4F{7/2} &    4097.26 &      0.735 &      0.731 \\

G[4]$^{\rm o}_{9/2}$ &          - &       \SLJ2F{7/2} &    4371.62 &      0.125 &      0.118 \\

G[4]$^{\rm o}_{9/2}$ &          - &       \SLJ4D{7/2} &    4332.70 &      0.121 &      0.133 \\

G[4]$^{\rm o}_{7/2}$ &          - &       \SLJ4F{7/2} &    4096.94 &      0.046 &      0.044 \\

G[4]$^{\rm o}_{7/2}$ &          - &       \SLJ4F{5/2} &    4083.90 &      0.428 &      0.423 \\

G[4]$^{\rm o}_{7/2}$ &          - &       \SLJ2F{5/2} &    4340.33 &      0.354 &      0.320 \\

G[4]$^{\rm o}_{7/2}$ &          - &       \SLJ4D{5/2} &    4331.17 &      0.123 &      0.159 \\

G[4]$^{\rm o}_{7/2}$ &          - &       \SLJ2D{5/2} &    4677.07 &      0.046 &      0.052 \\

G[3]$^{\rm o}_{7/2}$ &          - &       \SLJ4F{7/2} &    4108.76 &      0.054 &      0.054 \\

G[3]$^{\rm o}_{7/2}$ &          - &       \SLJ4D{7/2} &    4345.56 &      0.050 &      0.052 \\

G[3]$^{\rm o}_{7/2}$ &          - &       \SLJ4F{5/2} &    4095.64 &      0.296 &      0.280 \\

G[3]$^{\rm o}_{7/2}$ &          - &       \SLJ2F{5/2} &    4353.59 &      0.162 &      0.117 \\

G[3]$^{\rm o}_{7/2}$ &          - &       \SLJ4D{5/2} &    4344.38 &      0.189 &      0.221 \\

G[3]$^{\rm o}_{7/2}$ &          - &       \SLJ4P{5/2} &    4291.26 &      0.246 &      0.273 \\

F[4]$^{\rm o}_{9/2}$ &          - &       \SLJ4F{9/2} &    4062.93 &      0.150 &      0.152 \\

F[4]$^{\rm o}_{9/2}$ &          - &       \SLJ4F{7/2} &    4046.12 &      0.015 &      0.019 \\

F[4]$^{\rm o}_{9/2}$ &          - &       \SLJ2F{7/2} &    4313.44 &      0.153 &      0.166 \\

F[4]$^{\rm o}_{9/2}$ &          - &       \SLJ4D{7/2} &    4275.55 &      0.681 &      0.662 \\

F[4]$^{\rm o}_{7/2}$ &          - &       \SLJ4F{7/2} &    4044.95 &      0.024 &      0.022 \\

F[4]$^{\rm o}_{7/2}$ &          - &       \SLJ2F{7/2} &    4312.11 &      0.098 &      0.103 \\

F[4]$^{\rm o}_{7/2}$ &          - &       \SLJ4D{7/2} &    4274.25 &      0.035 &      0.030 \\

F[4]$^{\rm o}_{7/2}$ &          - &       \SLJ4F{5/2} &    4032.24 &      0.017 &      0.019 \\

F[4]$^{\rm o}_{7/2}$ &          - &       \SLJ4D{5/2} &    4273.10 &      0.081 &      0.070 \\

F[4]$^{\rm o}_{7/2}$ &          - &       \SLJ2D{5/2} &    4609.43 &      0.725 &      0.733 \\

F[4]$^{\rm o}_{7/2}$ &          - &       \SLJ4P{5/2} &    4221.70 &      0.017 &      0.020 \\

F[3]$^{\rm o}_{7/2}$ &          - &       \SLJ4F{9/2} &    4065.05 &      0.016 &      0.016 \\

F[3]$^{\rm o}_{7/2}$ &          - &       \SLJ4F{7/2} &    4048.22 &      0.094 &      0.099 \\

F[3]$^{\rm o}_{7/2}$ &          - &       \SLJ2F{7/2} &    4315.83 &      0.046 &      0.044 \\

F[3]$^{\rm o}_{7/2}$ &          - &       \SLJ4D{7/2} &    4277.90 &      0.154 &      0.159 \\

F[3]$^{\rm o}_{7/2}$ &          - &       \SLJ2F{5/2} &    4285.68 &      0.298 &      0.354 \\

F[3]$^{\rm o}_{7/2}$ &          - &       \SLJ4D{5/2} &    4276.75 &      0.314 &      0.265 \\

F[3]$^{\rm o}_{7/2}$ &          - &       \SLJ2D{5/2} &    4613.68 &      0.074 &      0.059 \\

D[3]$^{\rm o}_{7/2}$ &          - &       \SLJ4D{7/2} &    4358.44 &      0.026 &      0.021 \\

D[3]$^{\rm o}_{7/2}$ &          - &       \SLJ4F{5/2} &    4107.09 &      0.173 &      0.190 \\

D[3]$^{\rm o}_{7/2}$ &          - &       \SLJ2F{5/2} &    4366.53 &      0.050 &      0.047 \\

D[3]$^{\rm o}_{7/2}$ &          - &       \SLJ4D{5/2} &    4357.25 &      0.079 &      0.091 \\

D[3]$^{\rm o}_{7/2}$ &          - &       \SLJ2D{5/2} &    4707.50 &      0.014 &      0.016 \\

D[3]$^{\rm o}_{7/2}$ &          - &       \SLJ4P{5/2} &    4303.82 &      0.652 &      0.630 \\

\hline
\end{tabular}
\end{center}
\end{table}


\begin{table}
\begin{centering}
\caption{The intensity ratios $I_{4649}/I_{\mathrm{sum}}$ and the electron densities $N_{e}$(O\II)
as obtained from the present work for the observational data of Figure \ref{PeimbertP2005T10000K4d}
where in the $2^{nd}$ column H stands for H\II\ regions, PL for PNe with low ADF and PH for PNe
with high ADF. The line intensities, [Cl~{\sc iii}] densities and [O~{\sc iii}] temperatures are
taken from the cited references. The electron densities $N_{e}$ in the $4^{th}$ and $5^{th}$
columns are in $\mathrm{cm}{}^{-3}$.\vspace{0.2cm}\label{tabObjects}}
\begin{tabular*}{16cm}{@{\extracolsep{\fill}}llllllc}
\hline
 {\bf Object}  & {\bf Type} & {\boldmath $I_{4649}/I_{\mathrm{sum}}$} & {\boldmath $N_{e}({\rm O\,II})$} & {\boldmath $N_{e}([{\rm Cl\,III}])$} & {\CR{\boldmath $T_{e}([{\rm O\,III}])$}} & {\bf Reference}\tabularnewline %
 \hline 30 Doradus  & H  & 0.160$\pm$0.017 & 430$^{+110}_{-90}$ & 270$_{-230}^{+250}$ & {\CR 9950$\pm$60} & 1\tabularnewline
 30 Doradus  & H  & 0.174$\pm$0.038 & 570$_{-190}^{+280}$ & 480 & {\CR 10100} & 2\tabularnewline
 M42  & H    & 0.345$\pm$0.003  & 7480$^{+1600}_{-1320}$  &  & {\CR 9300} & 3\tabularnewline
 M42  & H    & 0.323$\pm$0.013 & 4660$_{-760}^{+900}$  & 9400$_{-700}^{+1200}$  & {\CR 8300$\pm$40} & 4\tabularnewline
 NGC 3576    & H & 0.258$\pm$0.015  & 1950$_{-270}^{+310}$  & 3500$_{-700}^{+900}$  & {\CR 8500$\pm$50} & 5\tabularnewline
 S311        & H & 0.180$\pm$0.070  & 540$_{-290}^{+620}$  & 550$_{-550}^{+350}$  & {\CR 9000$\pm$200} & 6\tabularnewline
 NGC 604     & H & 0.151$\pm$0.035  & 310$_{-140}^{+254}$ & 100$_{-50}^{+100}$  & {\CR 8150$\pm$160} & 7\tabularnewline
 NGC 5315    & PL & 0.373$\pm$0.025  & 10840$_{-1080}^{+8130}$  & 22825 & {\CR 8850} & 8\tabularnewline
 NGC 3132  & PL & 0.213$\pm$0.024 & 1160$_{-230}^{+290}$  & 720 & {\CR 9530} & 9\tabularnewline
 NGC 3242 & PL & 0.283$\pm$0.040 & 2590$_{-530}^{+670}$  & 1200 & {\CR 11700} & 9\tabularnewline
 NGC 3918 & PL & 0.290$\pm$0.021 & 3270$_{-700}^{+880}$ & 5500 & {\CR 12600} & 9\tabularnewline
 NGC 2022 & PH & 0.350$\pm$0.022  & 7450$_{-210}^{+2950}$  & 850 & {\CR 15000} & 9\tabularnewline
 NGC 6153  & PH & 0.333 & 12710$_{-5300}^{+9070}$  & 3830 & {\CR 9110} & 10\tabularnewline
 M1-42  & PH & 0.241$\pm$0.007 & 2810$_{-620}^{+800}$  & 1580$_{-460}^{+200}$  & {\CR 8900$\pm$300} & 11\tabularnewline
 NGC 7009 & PH & 0.349$\pm$0.017 & 10840$_{-4440}^{+7530}$ & 3600 & {\CR 9810} & 12\tabularnewline \hline
\end{tabular*}
\par\end{centering}

{\CR %
References:
 (1) \citet{Peimbert2003}
 (2) \citet{TsamisBLDS2003}
 (3) \citet{BaldwinFMCCPS1991, BaldwinVVFMe2000}
 (4) \citet{EstebanPGRPR2004}
 (5) \citet{RojasEPRRP2004}
 (6) \citet{RojasEPPRR2005}
 (7) \citet{EstebanBPGPM2009}
 (8) \citet{RuizPPE2003}
 (9) \citet{TsamisBLDS2003b}
 (10) \citet{LiuSBDCB2000}
 (11) \citet{McnabbFL2016}
 (12) \citet{FangL2011}.
}
\end{table}

\end{document}